\documentclass[letterpaper]{article} 
\usepackage{aaai2026}  
\usepackage{times}  
\usepackage{helvet}  
\usepackage{courier}  
\usepackage[hyphens]{url}  
\usepackage{graphicx} 
\usepackage{natbib}  
\usepackage{caption} 
\usepackage{algorithm}
\usepackage{algorithmic}

\usepackage{booktabs}
\usepackage{amsmath}
\usepackage{multirow}
\usepackage{amssymb}
\usepackage{newfloat}
\usepackage{listings}
\DeclareCaptionStyle{ruled}{labelfont=normalfont,labelsep=colon,strut=off} 
\floatstyle{ruled}
\newfloat{listing}{tb}{lst}{}
\floatname{listing}{Listing}
\title{MindVote: When AI Meets the Wild West of Social Media Opinion}
\author {
    Xutao Mao\textsuperscript{\rm 1},
    Ezra Xuanru Tao\textsuperscript{\rm 1},
    Leyao Wang\textsuperscript{\rm 2}
}
\affiliations {
    \textsuperscript{\rm 1}Vanderbilt University\\
    \textsuperscript{\rm 2}Yale University\\
    xutao.mao@vanderbilt.edu, 
    ezra.tao@vanderbilt.edu, 
    leyao.wang.lw855@yale.edu
}

\usepackage{bibentry}

\begin{document}

\maketitle

\begin{abstract}
Large Language Models (LLMs) are increasingly used as scalable tools for pilot testing, predicting public opinion distributions before deploying costly surveys. However, the prevailing paradigm for evaluating these models relies on traditional structured surveys—a methodology misaligned with the more realistic scenarios like social media where opinions are rich in digital contexts. By design, surveys strip away the social and cultural context that shapes public opinion, and LLM benchmarks built on this paradigm inherit these critical limitations. To bridge this gap, we introduce MindVote, the first benchmark for public opinion prediction grounded in authentic social media discourse. MindVote is constructed from 3,918 naturalistic polls sourced from Reddit and Weibo, spanning 23 topics and enriched with detailed annotations for platform and topical context. Using this benchmark, we conduct a comprehensive evaluation of 15 LLMs, revealing a critical ``\textit{survey-based specialization pitfall}" where models fine-tuned on traditional surveys underperform their general-purpose counterparts and demonstrating the necessity of context in social media. MindVote provides a robust, ecologically valid framework to move beyond survey-based evaluations and advance the development of more socially intelligent AI systems.
\end{abstract}

\section{Introduction}
\begin{figure}[t]
    \centering
    \includegraphics[width=1\linewidth]{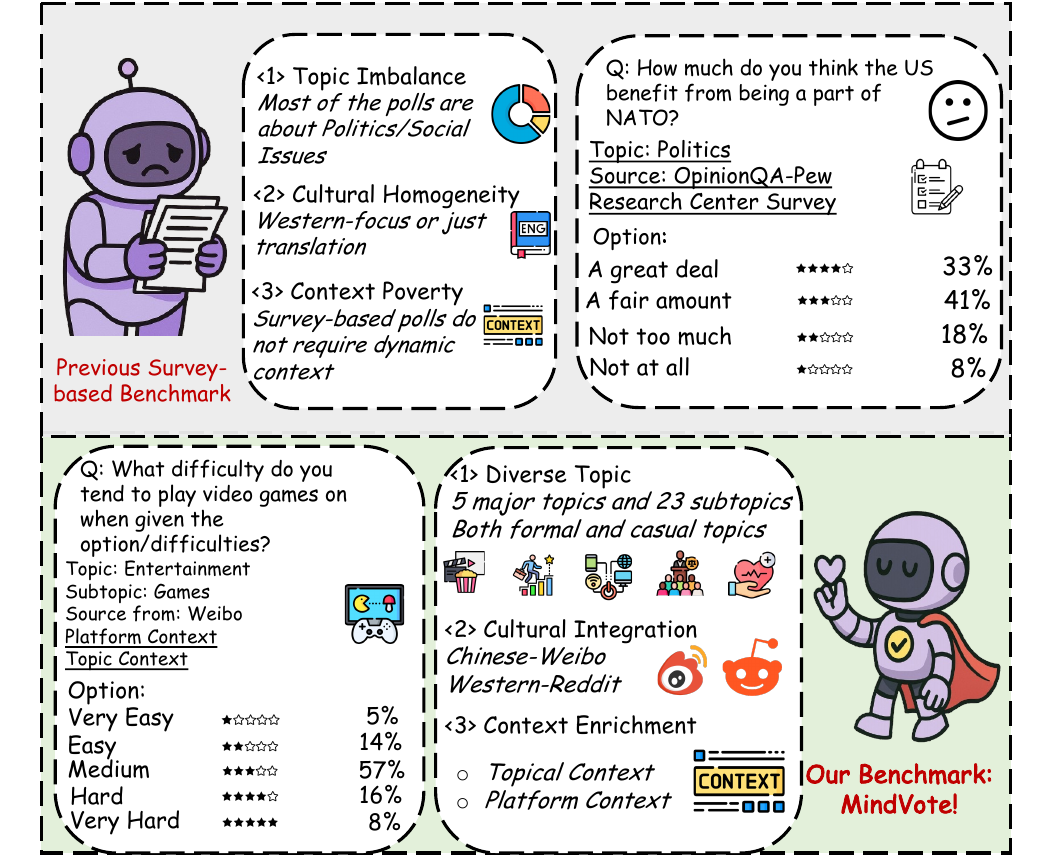} 
    \caption{MindVote benchmark addresses three key limitations of previous survey-based approaches. Our benchmark provides diverse topics, cultural integration, and rich contextual metadata, overcoming the topic imbalance, cultural homogeneity, and context poverty of traditional survey datasets.}
    \label{fig:overall}
\end{figure}

A core application for Large Language Models (LLMs) in understanding public opinion is serving as a rapid, scalable tool for pilot testing—predicting how a population will respond to a question before deploying costly, large-scale surveys \cite{rothschild2024opportunities,bisbee2024synthetic, suh2025language,sinacola2025llms,cao2025specializing,anthis2025llm,binz2025should}. This capability is crucial across a wide spectrum of domains. For example, LLMs can be used to predict reactions to new entertainment content and technological innovations, understand career aspirations, navigate complex social issues, and track evolving lifestyle trends \cite{argyle2023out,qian2025large,chen2025large}. However, the central goal is not simply to identify the majority opinion, but to predict how opinions are distributed across all possible choices. Understanding the full distribution is important because it reveals whether society is unified, divided, or contains strong dissent—information that a single majority number cannot capture. \cite{zhou-etal-2022-distributed,moon-etal-2024-virtual,meister-etal-2025-benchmarking,cao-etal-2025-specializing}. 

However, the prevailing paradigm for public opinion distribution prediction relies on traditional structured surveys \cite{krogstad2016pew,santurkar2023whose,elkjaer2024estimating,suh2025language}. This approach does not align with the fast-paced, context-rich digital environments—such as social media platforms—where opinions are now actively formed and expressed  \citep{boutyline2021cultural, riva2022social}. While surveys include basic demographic questions—such as age, gender, or education level—these static variables are poor proxies for the dynamic, situated nature of social context \cite{zaller1992simple}. Demographics may indicate who a person is in broad categories, but fail to reflect how opinions are specifically shaped by lived experiences, group affiliations, ongoing conversations, or culturally reasoning patterns \cite{turner1979social,kahan2011cultural,markus2014culture,mcintosh2019exploring}.

Consequently, LLM benchmarks built on this flawed paradigm, such as OpinionQA and SubPop \citep{santurkar2023whose, suh2025language}, inherit these critical limitations. This reveals a crucial gap in the responsible deployment of LLMs. As these models are increasingly used to act as a proxy for human populations, ensuring their accuracy becomes essential. A proper evaluation framework must identify when models fail at context-rich, real-world tasks. Without this, we risk promoting systems that misrepresent the groups they are meant to simulate. This misrepresentation has severe downstream consequences, potentially compromising strategic decisions across critical topics like public policy, marketing, technology development, and community management \citep{qi2024impact, radivojevic2024llms, qu2024performance, panickssery2024llm, repec:arx:papers:2502.03158, wang2025large}. This reliance on traditional survey-based evaluation creates three critical gaps in our ability to assess an LLM's true social understanding:

\noindent \textbf{The Gap about Topic Imbalance.} Real-world online discourse is thematically diverse \cite{tran-ostendorf-2016-characterizing,song2023understanding}; 
for instance,  entertainment drives 70\% of traffic on Weibo \citep{chinamarketingcorp2024weibo}, while Reddit's largest communities focus on gaming and technology \citep{proferes2021studying}.
Yet, current survey-based benchmarks are heavily skewed towards formal, institutional topics like politics, which represent only a fraction of naturalistic opinion expression \citep{santurkar2023whose,zhao-etal-2024-worldvaluesbench}. Because of this mismatch, we cannot evaluate if a model can adjust its reasoning from formal political surveys to the everyday language used on social media \cite{karjus2024evolving,reveilhac2024augmenting}.

\noindent \textbf{The Gap about Cultural Homogeneity.} Existing cross-cultural survey benchmarks often exhibit cultural homogeneity, creating test conditions on Western-centric questions even with other languages translation \citep{haerpfer2022world, durmus2024towards, zhao-etal-2024-worldvaluesbench}. A model might perform well on a translated question, but we are left unable to determine if this success stems from genuine cultural understanding or a superficial linguistic competence \cite{singh-etal-2025-global}. 

\noindent\textbf{The Gap about Context Poverty.} Opinions on social media are not noise but are helpful predictive signals \cite{lopez2021framing}. This is correlated with \textit{context priming}, a phenomenon where exposure to a specific context influences how individuals perceive and respond \cite{doyle2016context}. On social platforms, users are primed by factors such as platform norms (e.g., Reddit's anonymity vs. Weibo's public-facing profiles \citep{brown2018reddit,zhu2024privacy}), temporal events \cite{banducci2015surveys}, and community-specific discourse \citep{tan2016winning, qiao2023does}. In contrast, survey-based benchmarks systematically remove this contextual priming to achieve standardization \citep{tourangeau2003context}, which prevents a true evaluation of a model's ability to leverage these critical predictive cues \cite{10.24963/ijcai.2023/749}.

To overcome these limitations, we move beyond surveys and instead use naturalistic social media polls—a more direct source for analyzing public opinion formation \cite{scarano2024analyzing}. These polls address three critical gaps and provide a robust foundation for evaluation. The comparison between existing survey-based benchmarks and our benchmark is in Figure \ref{fig:overall}.

\begin{figure}
    \centering
    \includegraphics[width=\linewidth]{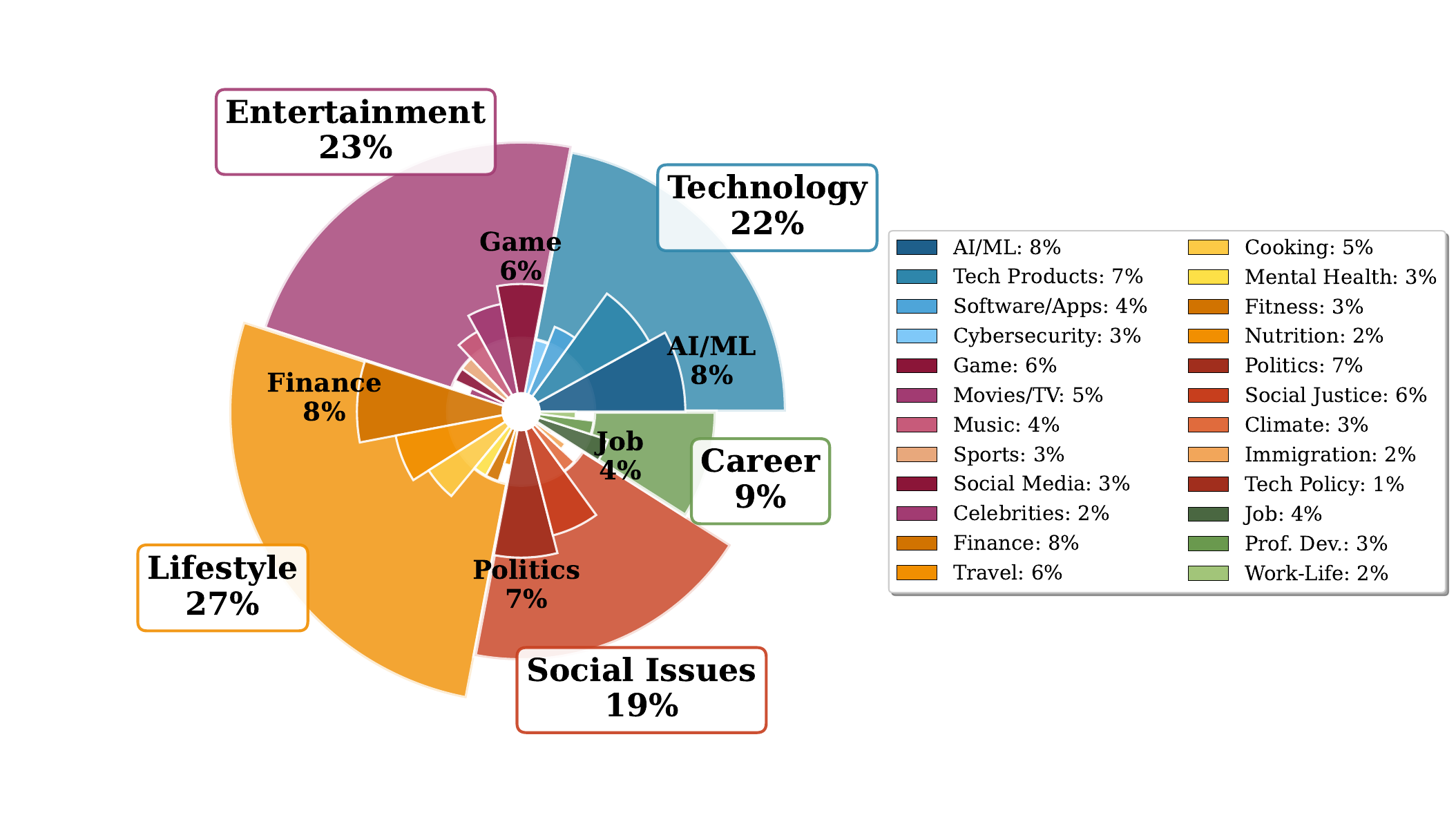}
    \caption{Distribution of topics in the MindVote dataset across five major topics and 23 subtopics. Percentages indicate the relative frequency of each subtopic.}
    \label{fig:topic}
\end{figure}
\noindent \textbf{Our Contributions.} We introduce \textbf{MindVote}, the first benchmark for public opinion distribution prediction grounded in realistic social media discourse. We advance the field with the following innovations:
\begin{itemize}
    \item We construct and release MindVote, a benchmark of 3,918 authentic polls from two platforms (Reddit and Weibo) in their native English and Chinese. It spans 5 major topics and 23 sub-topics (shown in Figure \ref{fig:topic}) and is enriched with detailed social context annotations. 
    \item We benchmark leading LLMs, identify top performers, and reveal that models fine-tuned on traditional surveys face a critical \textit{survey-based specialization pitfall}.
    \item We demonstrate that enhancing a model's capacity with social-context reasoning is more effective than fine-tuning on context-stripped data, offering a new direction for developing more socially intelligent systems.
\end{itemize}

\section{Related Work}
\noindent \textbf{Opinion Distribution Prediction Benchmarks.}
U.S.-focused benchmarks like OpinionQA \citep{santurkar2023whose} and SubPop \citep{suh2025language} concentrate on political topics, while international efforts like GlobalOpinionQA \citep{durmus2024towards} is Western-centric frameworks and WorldValuesBench \citep{zhao-etal-2024-worldvaluesbench} is framed as multi-cultural value prediction. \citet{cao2025specializing} introduces three datasets (two English, one Chinese) specifically for group-level distribution prediction. These approaches contain only demographic-value pairings, structured survey format, or Western-centric lens
that abstract away the naturalistic cultural contexts that are essential for authentic opinion distribution prediction.

\begin{figure}[htbp]
    \centering
    \includegraphics[width=\linewidth]{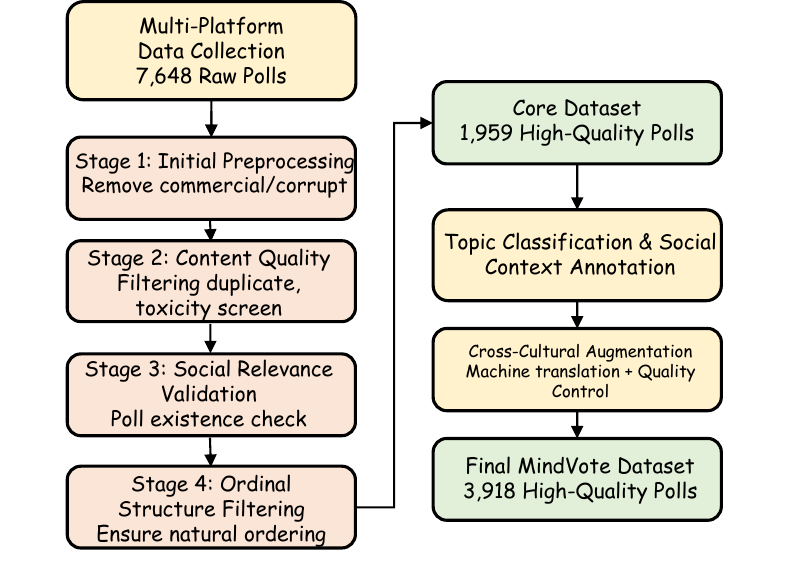}
    \caption{This flow pipeline demonstrates the dataset construction from data source to final composition of datasets.}
    \label{fig:construction}
\end{figure}

\section{Benchmarking Setup}
\subsection{Dataset Construction}

\label{sec:dataset_construction}

MindVote's construction involved transforming 7,648 raw polls into 3,918 high-quality polls through strategic data sourcing, social context annotation, rigorous quality control, and the cross-cultural augmentation. Figure \ref{fig:construction} shows the flow of dataset construction.

\noindent\textbf{Platform Selection Strategy.} Our platform selection strategy is designed to evaluate distinct aspects of LLM opinion prediction capabilities. Reddit provides an anonymous environment where users express unfiltered, authentic opinions without identity-based social constraints, requiring models to interpret and predict genuine thoughts purely through content analysis without relying on user profiles or reputation cues. Weibo enables the evaluation of model performance in culture-specific contexts, testing models' understanding of Chinese cultural contexts, social norms, and culture-laden discourse patterns.

\noindent\textbf{Multi-Platform Data Collection.} We collected 7,648 polls across two platforms spanning 2019-2025. Weibo dataset contributed 3,757 polls (2,026 from existing datasets \cite{lu-etal-2021-engage} from 2019 to 2021, 1,734 newly crawled) and are anonymized for users' personal identifiable information, capturing Chinese social media dynamics across pandemic and post-pandemic periods. Reddit provided 3,891 polls from diverse subreddits including \textit{r/poll} during 2021-2025.

\noindent\textbf{Quality Control Pipeline.} Our four-stage pipeline efficiently produced high-quality 1,959 core polls:

\begin{enumerate}
 \item \textbf{Initial Preprocessing:}
This process included the removal of commercial votes, targeting promotional polling content that lacked authentic user engagement. Furthermore, format-corrupted poll data was identified and discarded through a systematic validation of structural integrity. 
 \item \textbf{Content Quality Filtering:}
This filtering begins with duplicate content removal using a MinHash algorithm to eliminate items with greater than 95\% overlap \citep{shrivastava2014defense}, effectively targeting redundant trending topics, reposted polls, and cross-platform duplicates. Subsequently, automated toxicity screening was conducted using the Google Perspective API, with polls retaining a toxicity score below 0.4 being retained to filter out hate speech and overly controversial subjects \citep{lees2022new}. 

 \item \textbf{Social Relevance Validation:}
 Human verification of voting patterns (removing polls $\leq$ 100 votes for social relevance) and verification of poll and vote existence in all two platforms to ensure authentic social engagement and meaningful community participation. 

 \item \textbf{Ordinal Structure Filtering:}
 We performed systematic filtering to ensure all selected polls exhibit natural ordering relationships between options to ensure structural alignments with existing benchmarks \cite{santurkar2023whose}.  We systematically excluded polls with purely categorical options (e.g., preference choices among unordered alternatives). This filtering employed LLM-as-a-judge (DeepSeek-R1 \citep{2501.12948}), validated through human annotation achieving Fleiss'$\kappa$ = 0.59 agreement. 
 \end{enumerate}

\noindent\textbf{Topic Classification and Social Context Annotation.} Unlike survey-based benchmarks that focus on pan-political and social topics, MindVote includes 5 major topics and 23 specific topics. We assigned topic labels through a classification process combining automated judgment and human validation. Initial classification used content-based detection employing LLM-as-a-judge (Deepseek-R1) to identify primary topic. Human validation was applied selectively to ambiguous cases where automated classification was uncertain or polls exhibited mixed topic characteristics \cite{chen-etal-2024-humans}. Trained annotators using standardized topic definitions achieved inter-annotator agreement of Fleiss'$\kappa$ = 0.55 for these edge cases, with expert consensus resolving conflicts and assigning polls spanning multiple topics to their primary thematic focus. 

We manually annotate each poll includes rich metadata for social context enrichment: general platform context (e.g., user statistics and user behaviors) and topical context (topic-specific discourse patterns).

\noindent\textbf{Cross-Cultural Augmentation.} We created a parallel bilingual corpus for all two platforms through machine translation with rigorous quality control to demonstrate both linguistic and cultural effects: 10\% back-translation validation (BLEU $>$ 35 \citep{papineni2002bleu}), 5\% native speaker rewriting, and expert review (Fleiss' $\kappa$ = 0.51). The total number of polls becomes 3,918 after augmentation.

\noindent\textbf{Final Dataset Composition.} The final MindVote dataset is composed of 3,918 polls, with 2,158 sourced from Reddit and 1,760 from Weibo. Each poll is enriched with a comprehensive set of metadata, including its creation time, total vote count, and layers of social context: general platform context and topical context within that platform. To ensure broad accessibility and ease of use, the entire dataset is provided in both CSV and JSON formats. Table \ref{tab:poll_examples} shows an example with its simplified metadata keywords.

\begin{table}[t]
\centering
\small 

\begin{tabular}{@{} l p{5.5cm} @{}} 
\toprule
\multicolumn{2}{@{}l@{}}{\textbf{Poll: How threatened do you feel by AI replacing your job?}} \\
\multicolumn{2}{@{}l@{}}{\small(Platform: Reddit, Date: April 14, 2025, Votes: 6,567)} \\
\midrule
\textbf{Options} & 
    \begin{enumerate}
        \item Not at all threatened
        \item Slightly threatened
        \item Moderately threatened
        \item Very threatened
    \end{enumerate} \\
\midrule

\textbf{Platform Context} & Reddit user base: 58\% US, 46\% college-educated, generally tech-oriented. \\
\textbf{Topical Context} & Broad AI adoption. 78\% of organizations use AI; 55\% of Americans use AI regularly. \\
\bottomrule
\end{tabular}
\caption{An example poll from our MindVote dataset, demonstrating the structure of the question, options, and associated social context provided for model evaluation.}
\label{tab:poll_examples}
\end{table}

\subsection{Experiment Design}

\noindent\textbf{Task.} We evaluate LLMs on opinion distribution prediction within naturalistic social discourse in MindVote dataset including 3,918 polls. Given a poll question with metadata including social contexts, models predict the probability distribution over answer choices.

\noindent\textbf{Models.} We evaluate 15 leading LLMs across closed-source, open-source and specialization categories: Claude-3.7-Sonnet-02-24 \citep{claude-3.7-thinking}, GPT-4o-2024-11-20 \citep{openai-4o}, GPT-4.1-04-14 \citep{openai-4.1}, Gemini-2.5-Pro-05-06 \citep{google-gemini-2-5-pro}, o3-medium-04-16 \citep{openai-o3-o4mini-system-card}, Deepseek-R1-05-28 \citep{2501.12948}, Qwen2.5-32B-Instruct \citep{qwen2.5}, Gemma-2-9B-it \citep{team2024gemma}, Mistral-7B-Instruct-v0.1 \cite{chaplot2023albert}, Llama-2-13B-Chat \cite{touvron2023llama}, Llama-3-70B-Instruct \citep{grattafiori2024llama}, and Llama-4-Maverick-17B-128E \citep{llama4}. We also evaluate three opinion distribution prediction specialization models from \citet{suh2025language}. These models are fine-tuned using LoRA \cite{hu2022lora} on base models: Llama-3-70B-Base \cite{grattafiori2024llama}, Llama-2-13B-Base \cite{touvron2023llama}, and Mistral-7B-v0.1 \cite{chaplot2023albert} which we named them respectively as: SubPop-Llama-3-70B, SubPop-Llama-2-13B, SubPop-Mistral-7B. 

\noindent\textbf{Pipeline.} All primary evaluations use a greedy decoding strategy (temperature=0) with default hyperparameter settings under zero-shot with context annotation \cite{kojima2022large,han2025zerotuning}. For the specialization fine-tuned models, we adapt those models into our pipeline by first loading the respective pretrained base model and then applying the publicly available LoRA weight checkpoints provided by \citet{suh2025language}. 

\noindent\textbf{Prompt.} To ensure consistent and machine-readable outputs, we employ a structured prompting strategy where the model is given a JSON object containing the poll and its context. The template instructs the model to assume the role of a \textit{``opinion distribution prediction expert analyzing voting patterns and social dynamics."} The prompt includes the poll question and is enriched with social context metadata, with instruction for step-by-step reasoning. The model's task is to return a JSON object with a schema identical to the input, replacing placeholder fields with its numeric predictions for the voting distribution. 

\noindent\textbf{Evaluation Metrics.} We adopt four distinct metrics to provide a comprehensive evaluation. Our primary metric is \textbf{1 - Wasserstein Distance (1-Wass.)} \citep{santurkar2023whose, meister-etal-2025-benchmarking,suh2025language}. The Wasserstein Distance measures the minimum cost for transforming one distribution into another, crucially accounting for semantic similarity between answer choices by treating them as points in a metric space. To complement this, we also report \textbf{Spearman's Rank Correlation Coefficient ($\rho$)} \citep{zhou-etal-2022-distributed,pavlopoulos2023distance}, a non-parametric measure of how well the predicted ranking of options matches the true ranking of vote shares; \textbf{1 - KL Divergence} \citep{meister-etal-2025-benchmarking,nguyen2025task}, which quantifies the information loss when using the model's predicted distribution to approximate the ground truth; and \textbf{One-hot Accuracy} \citep{zhou-etal-2022-distributed,santurkar2023whose,suh2025language}, which provides a strict measure of whether the single most likely predicted answer is correct.

\noindent\textbf{Evaluation Boundary.}
We include upper bounds and lower bounds for comparisons following \cite{suh2025language}. The upper bound is established by sampling subsets of the original results, calculating the four metrics between subsampled and original distributions, and performing bootstrapping to obtain a robust estimate that captures the intrinsic variance arising from the respondent sampling process in opinion. The uniform distribution lower bound establishes a performance floor equivalent to random chance.

\section{Results}
\label{sec:main}

\subsection{Performance of General Purpose Models}
We analyze the overall performance of ten leading general purpose LLMs on the MindVote benchmark in Table \ref{tab:overall_performance_decimal}. Across all four metrics, the closed-source model o3-medium consistently outperforms all other models, establishing the highest performance ceiling among current general-purpose LLMs. Among open-source models, Deepseek-R1 demonstrates the strongest results, narrowly outperforming other models in its class. However, a critical gap remains between the top-performing models and the upper bound for all metrics. This highlights substantial opportunities for improvement in modeling authentic public opinion distribution.

\subsection{Survey-based Specialization Pitfalls}
\label{sec:specialization}

Our results in Table \ref{tab:overall_performance_decimal} show that fine-tuning on a structured survey dataset leads to a significant decrease in performance when predicting opinion distributions on social media. To demonstrate this, we compare our specialization models (base models that fine-tuned on traditional but context-stripped survey data) against their original base instruction-tuned models (the general-purpose models).


As shown in Table \ref{tab:overall_performance_decimal}, specialization models consistently underperform their base counterparts, even when evaluated with full context annotations. For example, the SubPop-Llama-2-13B specialization model experiences a 3.3 percentage point decrease in its 1-Wass. score relative to the base Llama-2-13B. These results indicate that fine-tuning on sanitized data does not equip models to handle the complexities of real-world contexts. Rather than enhancing performance, such narrow specialization actually impairs the models’ ability to generalize to authentic social discourse.

\begin{table}[htbp]
\centering
\small
\setlength{\tabcolsep}{1mm}

\begin{tabular}{lcccc}
\toprule
\textbf{Model} & \textbf{1-Wass. } & \textbf{1-KL Div.} & \textbf{Spearman.} & \textbf{Acc.} \\
\midrule
\multicolumn{5}{l}{\textit{Closed-source Models}} \\
\textbf{o3-medium} & \textbf{0.892} & \textbf{0.859} & \textbf{0.756} & \textbf{0.581} \\
Gemini-2.5-Pro & 0.891 & 0.845 & 0.751 & 0.564 \\
Claude-3.7-Sonnet & 0.891 & 0.851 & 0.722 & 0.551 \\
GPT-4o & 0.880 & 0.836 & 0.691 & 0.515 \\
GPT-4.1 & 0.874 & 0.845 & 0.688 & 0.524 \\
\midrule
\multicolumn{5}{l}{\textit{Open-source Models}} \\
Deepseek-R1 & 0.876 & 0.831 & 0.739 & 0.558\\
Qwen2.5-32B & 0.866 & 0.787 & 0.605 & 0.483 \\
Llama-4-17B & 0.820 & 0.731 & 0.659 & 0.429 \\
Llama-3-70B & 0.844 & 0.752 & 0.641 & 0.461 \\
Llama-2-13B & 0.807 & 0.718 & 0.592 & 0.369 \\
Gemma-2-9B & 0.802 & 0.705 & 0.575 & 0.362 \\
Mistral-7B & 0.808 & 0.719 & 0.597 & 0.365 \\
\midrule
\multicolumn{4}{l}{\textit{Specialization Models}} \\
SubPop-Llama-3-70B & 0.805 & 0.713 & 0.593 & 0.417 \\
SubPop-Llama-2-13B & 0.774 &0.693 &  0.558 & 0.378 \\
SubPop-Mistral-7B & 0.782 &0.695  & 0.546 & 0.370\\
\midrule
Upper Bound & 0.972 & 0.976 & 0.961 & 0.964 \\
Lower Bound & 0.701 & 0.663 & 0.000 & 0.307\\
\bottomrule
\end{tabular}
\caption{Opinion distribution prediction performance of LLMs on the MindVote Benchmark. Scores are presented as Mean values, evaluated on four different metrics: 1-Wasserstein distance (1-Wass.), 1-KL Divergence (1-KL Div.), Spearman's Rank Correlation (Spearman.), and One-hot Accuracy (Acc.). \textbf{All metrics are the higher the better. }}
\label{tab:overall_performance_decimal}
\end{table}

\subsection{Performance Across Topics and Cultures}
\label{sec:performanceanalysis}
\noindent\textbf{Performance varies by Topic.} As shown in Table~\ref{tab:topic_performance}, model performance differs notably across topical domains. On average, models perform better on \textit{Social Issues} and \textit{Lifestyle} topics, which typically exhibit more structured discourse aligned with standard pre-training corpora. In contrast, performance declines in domains such as \textit{Technology} and \textit{Entertainment}, where specialized jargon, community-specific norms, and rapidly evolving vernacular are prevalent. This disparity indicates that models have difficulty adapting their reasoning from general knowledge to the nuanced linguistic and social norms of diverse online communities.

\begin{table}[htbp]
\centering
\small
\setlength{\tabcolsep}{1mm}

\begin{tabular}{lccccc}
\toprule
\textbf{Model} & \textbf{Tech.} & \textbf{Soc.} & \textbf{Enter.} & \textbf{Career} & \textbf{Life.} \\
\midrule
\multicolumn{6}{l}{\textit{Closed-source Models}} \\
o3-medium & 0.868 & \textbf{0.925} & 0.869 & \textbf{0.904} & 0.903 \\
Gemini-2.5-Pro & 0.860 & 0.917 & 0.881 & 0.896 & \textbf{0.906} \\
Claude-3.7-S. & 0.865 & 0.908 & \textbf{0.896} & 0.891 & 0.894 \\
GPT-4o & 0.859 & 0.893 & 0.872 & 0.895 & 0.884 \\
GPT-4.1 & 0.866 & 0.913 & 0.854 & 0.878 & 0.868 \\
\midrule
\multicolumn{6}{l}{\textit{Open-source Models}} \\
Deepseek-R1 & \textbf{0.889} & 0.892 & 0.868 & 0.879 & 0.887 \\
Qwen2.5-32B & 0.846 & 0.876 & 0.867 & 0.868 & 0.875 \\
Llama-4-17B & 0.824 & 0.844 & 0.832 & 0.866 & 0.857 \\
Llama-3-70B & 0.835 & 0.853 & 0.836 & 0.845 & 0.854 \\
Llama-2-13B & 0.798 & 0.824 & 0.801 & 0.789 & 0.817 \\
Gemma-2-9B & 0.802 & 0.839 & 0.832 & 0.848 & 0.885 \\
Mistral-7B & 0.809 & 0.836 & 0.812 & 0.804 & 0.828 \\
\midrule
\multicolumn{6}{l}{\textit{Specialization Models}} \\
SubPop-Llama-3-70B & 0.794 & 0.821 & 0.798 & 0.786 & 0.813 \\
SubPop-Llama-2-13B & 0.762 & 0.798 & 0.769 & 0.751 & 0.785 \\
SubPop-Mistral-7B & 0.771 & 0.803 & 0.778 & 0.764 & 0.792 \\
\midrule
\textbf{Average} & 0.830 & 0.863  & 0.838  & 0.844 & 0.857 \\
\bottomrule
\end{tabular}
\caption{Opinion distribution prediction performance across different topics. The final row shows the average performance across all models for each major topic.}
\label{tab:topic_performance}
\end{table}

\begin{figure}[htbp]
    \centering
    \includegraphics[width=\columnwidth]{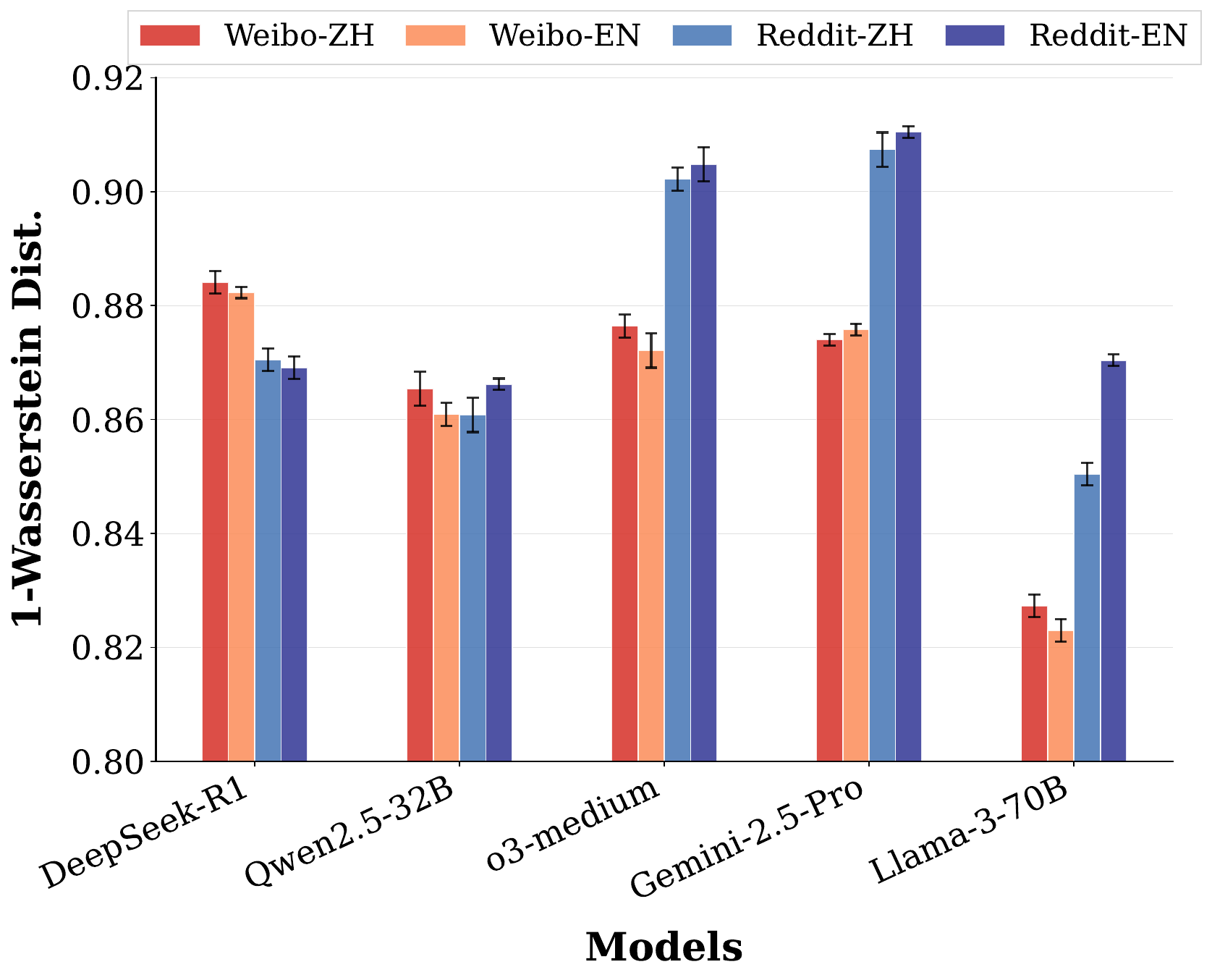} 
    \caption{LLMs exhibit a cultural alignment with their origin. Models trained from Western data better on Reddit (EN-English), while those from China (e.g., DeepSeek-R1) excel on Weibo (ZH-Chinese). This cultural gap persists after controlling for linguistic effects. Error bars represent 95\% CI.}
    \label{fig:cultural}
\end{figure}

\noindent\textbf{Performance Reflects Cultural Origin.} The analysis in Figure \ref{fig:cultural} reveals a \textit{Cultural Gap}, showing models have a strong cultural alignment with their origin. Western models like Gemini-2.5-Pro excel on Reddit, while Chinese models such as DeepSeek-R1 dominate on Weibo. We confirmed this gap is primarily cultural, not linguistic, by evaluating translated content; the performance penalty from translation was minimal compared to the large drop across cultural sources. This systematic variation demonstrates a clear alignment effect, where models are more adept at interpreting the cultural norms of their training data.

\subsection{Analysis}
\label{sec:analysis}

Section \ref{sec:specialization} highlights a pitfall of survey-based specialization: base models fine-tuned on sanitized survey datasets show degraded performance, even when \textit{context priming}—adding relevant situational or background information to guide predictions—is applied \cite{doyle2016context}.
This unexpected finding raises three questions that we expect to investigate: 

\begin{itemize}
    \item 
    Does social context in social media serve as noise or a helpful signal?
    \item How does increasing contextual complexity affect model performance, especially for survey-specialized models?
    \item Does adding social contexts better enhance models' opinion prediction compared with few-shot learning?
\end{itemize}

These questions emerge naturally from our specialization findings. 
The failure of fine-tuned models, even when provided with full context, suggests two possibilities: either (1) platform and topic are less important than previously assumed, or (2) these models are unable to effectively utilize such contextual information.
This motivates our first investigation into whether social context is truly beneficial signal or noise. Moreover, the suboptimal performance of specialized models reflects their difficulty in processing social context, motivating our second analysis on factors contributing to context complexity.
Third, because our context-rich approach relies on contextual understanding rather than example-based training as in few-shot learning, a direct comparison is needed to assess the effectiveness of both strategies.

\noindent\textbf{Social Context is a Critical Signal.} Our context ablation study (Table \ref{tab:context_ablation_drops_only}) confirms that social context is not noise but a vital signal for accurate opinion prediction. Removing all contextual metadata (No Context) results in the most substantial performance drop on average (\textit{-5.91\%}), followed closely by platform-specific information removal (\textit{-5.12\%}) and topical context removal (\textit{-4.52\%}). 

Notably, specialized fine-tuned models exhibit dramatically larger degradation across all ablation conditions. This severe degradation approaches the theoretical lower bound of uniform distribution performance, suggesting these models have become overly dependent on contextual signals used for priming. Since these models were fine-tuned on structured survey data in standardized formats, they struggle to adapt when deployed on natural social media polls when contextual information may be fragmented or entirely absent. While these models may excel in controlled survey environments, their heightened sensitivity to context removal in real-world settings reveals a limitation, constraining their applicability across diverse online environments.

\begin{table}[t]
\centering
\small
\begin{tabular}{lrrr}
\toprule
\textbf{Model} & \textbf{w/o Plat.} & \textbf{w/o Topic.} & \textbf{No Ctx.} \\
\midrule
\multicolumn{4}{l}{\textit{Closed-source Models}} \\
o3-medium & -3.03 & -2.31 & -4.13 \\
Gemini-2.5-Pro & -2.37 & -3.30 & -4.34 \\
Claude-3.7-Sonnet & -4.46 & -3.91 & -5.53 \\
GPT-4o & -5.75 & -3.39 & -4.92 \\
GPT-4.1 & -4.92 & -3.36 & -5.30 \\
\midrule
\multicolumn{4}{l}{\textit{Open-source Models}} \\
Deepseek-R1 & -4.19 & -3.55 & -5.98 \\
Qwen2.5-32B & -4.42 & -3.81 & -6.24 \\
Llama-3-70B & -4.82 & -4.78 & -5.19 \\
Llama-4-17B & -5.80 & -5.33 & -6.08 \\
Llama-2-13B & -4.95 & -5.26 & -5.92 \\
Gemma-2-9B & -5.37 & -4.74 & -6.47 \\
Mistral-7B & -5.92 & -4.84 & -6.55 \\
\midrule
\multicolumn{4}{l}{\textit{Specialization Models}} \\
SubPop-Llama-3-70B & -6.82 & -6.68 & -6.95 \\
SubPop-Llama-2-13B & -6.94 & -6.11 & -7.23 \\
SubPop-Mistral-7B & -6.89 & -6.41 & -7.54 \\
\midrule
\textbf{Average} & -5.12 & -4.52 & -5.91 \\
\bottomrule
\end{tabular}
\caption{Opinion distribution prediction performance degradation from the full-context baseline. All scores represent the drop in 1-Wasserstein Distance (\%). Abbreviations: Plat. (Platform), Topic. (topical), Ctx. (Context). }
\label{tab:context_ablation_drops_only}
\end{table}

\begin{figure}
    \centering
    \includegraphics[width=\linewidth]{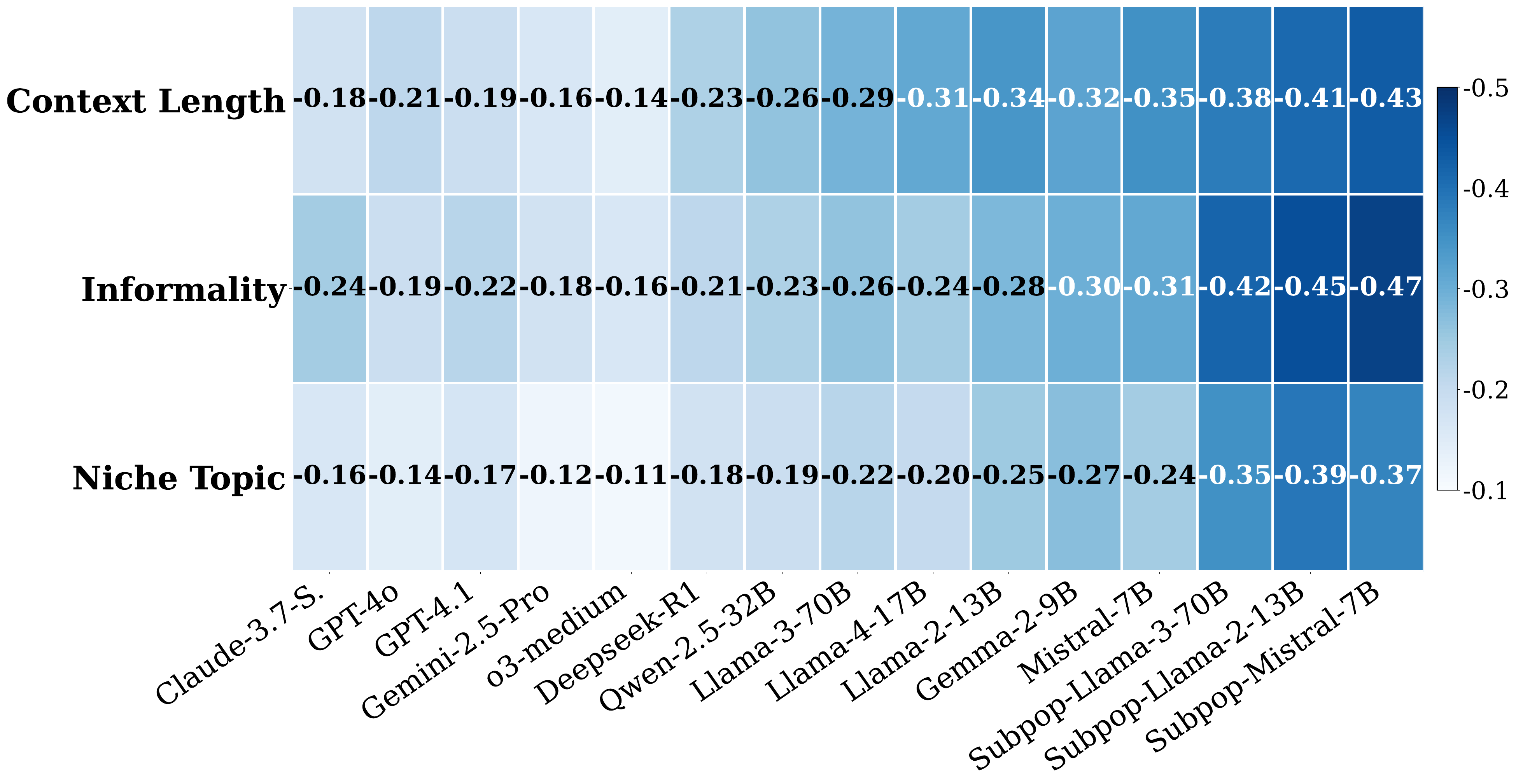}
    \caption{Correlation patterns between model's 1-Wass. performance and complexity dimensions. Survey-specialized models exhibit strong brittleness to social media contextual complexity. }
    \label{fig:heatmap}
\end{figure}

\noindent\textbf{Contextual Complexity.} While our ablation study confirms that social context is critical for performance, its inherent complexity presents a significant challenge. To dissect this, we investigate how model performance degrades as complexity increases about our annotated context. We define contextual complexity using three metrics: (1) context length by tokens of combined metadata; (2) language informality in the context; and (3) niche topic which constitutes a small fraction of the overall dataset. Our analysis in Figure \ref{fig:heatmap} reveals that increasing complexity in any of these areas has negative correlations with the 1-Wass. performance. 

This effect is most pronounced for our specialization models. Specifically, these models, fine-tuned on structured and formal survey data, exhibit strong performance degradation when faced with longer contexts, higher degrees of informality, and more niche topics. This demonstrates a critical brittleness: while trained to be domain experts, their reliance on formal data structures makes them particularly vulnerable to the unstructured, multifaceted, and messy nature of authentic social media discourse. 

\noindent\textbf{Contextual Priming vs. Few-Shot Learning.} Our ``survey-based specialization pitfall" finding (Section \ref{sec:specialization}) raised a key question: do models benefit more from understanding a poll's environment via contextual priming or from imitating examples via few-shot learning? To evaluate this, we compared the model's performance across three distinct settings. The first was a zero-shot setting with context priming, which followed our default context annotation pipeline. We contrasted this against two settings without context priming: a standard zero-shot setting without context priming, same as we have done in Table \ref{tab:context_ablation_drops_only} and a few-shot (1-4 examples) setting. For the few-shot configuration, we supplied the model with examples predicted from Claude-3.7, where each example contained a poll, its distribution prediction, and the corresponding reasoning steps.

\begin{figure}
    \centering
    \includegraphics[width=\linewidth]{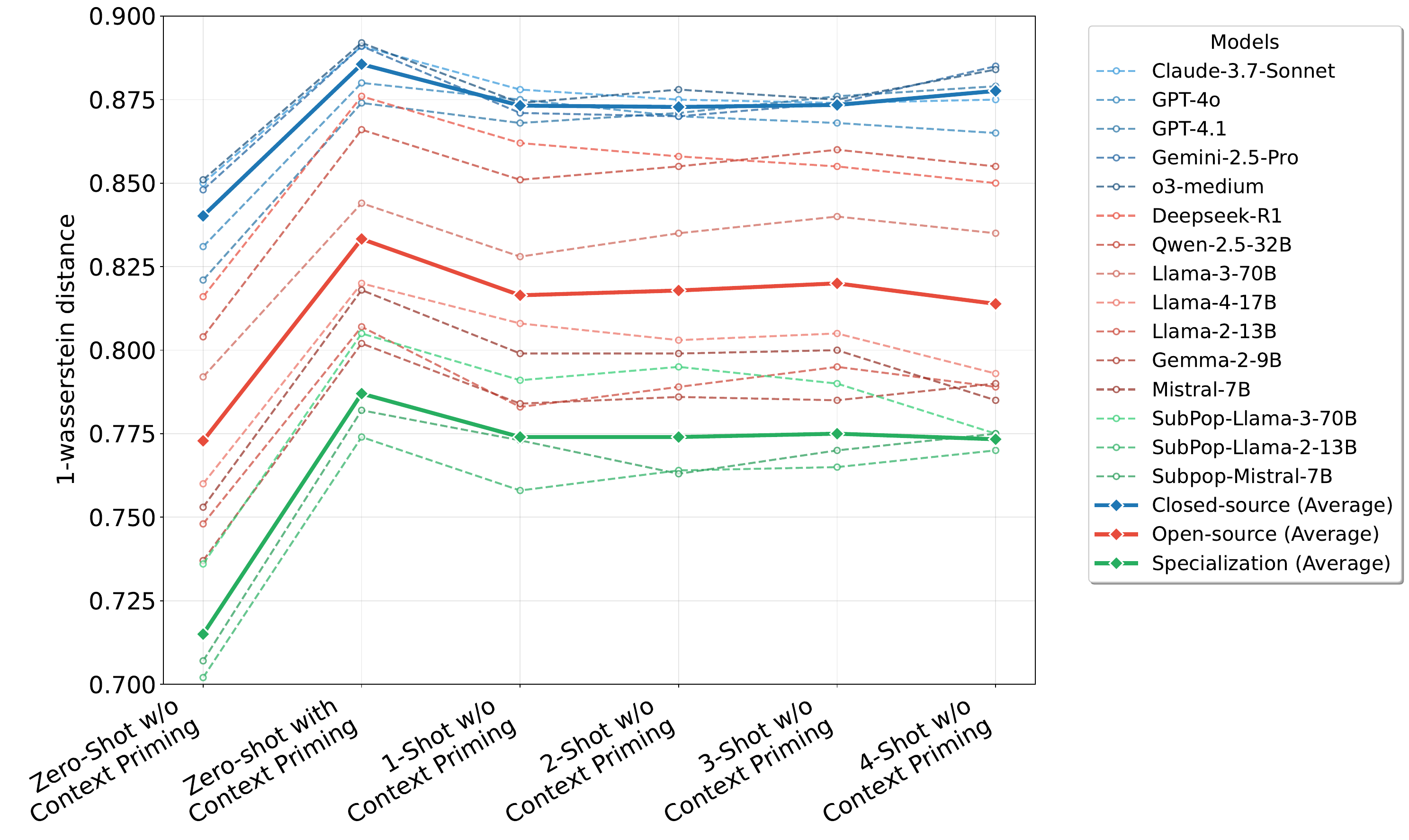}
    \caption{Contextual priming outperforms few-shot learning across all model categories, highlighting the importance of social context over example-based pattern matching.}
    \label{fig:context}
\end{figure}

The results in Figure \ref{fig:context} are stark: contextual priming delivers a substantial performance boost across all models, often exceeding the gains achieved through few-shot learning. In contrast, the benefit from few-shot examples is unreliable. As the figure shows, performance does not reliably increase with the number of examples; for instance, the average 4-shot performance of open-sourced models is worse than that of 3-shot performance. This discrepancy suggests that the ability to situate a problem within its broader social context is a more critical and robust capability than simple pattern-matching from isolated examples.

\section{Error Analysis}
We conduct an error analysis to categorize prediction reasoning errors \footnote{We consider the individual poll's 1-Wass. $<$ 0.8 as error.}. Using the LLM-as-a-judge, we identify three main categories of error type shown in Table \ref{tab:error_analysis}.

\begin{table}[t]
\centering
\small 
\setlength{\tabcolsep}{2mm} 

\begin{tabular}{@{}p{0.25\columnwidth}p{0.65\columnwidth}@{}}
\toprule
\textbf{Error Type} & \textbf{Case Study \& Failed Reasoning} \\
\midrule

\textbf{Platform Misadaptation} 
& \textit{Influencer Subscription Worth or Not (Reddit):} Failing to recognize Reddit attitude toward ad-centric monetization and user engagement patterns. \\

\textbf{Cultural Misalignment} 
& \textit{New phone Buy or Wait (Weibo):} Assumed universal tech-enthusiasm drives upgrades, overlooking Chinese market's specific socio-economic factors and consumer behavior patterns. \\

\textbf{Temporal Dislocation} 
& \textit{Calling it "X" vs. "Twitter" (Reddit):} Reasoning anchored to official rebrand timeline, underestimating persistent colloquial usage and real-world adoption resistance. \\

\bottomrule
\end{tabular}
\caption{Analysis of Claude 3.7 Sonnet prediction errors categorized by failure modes: \textbf{Platform Mis-Adaptation} (42.5\%) -- misapplying knowledge about specific platforms; \textbf{Cultural Misalignment} (36.6\%) -- ignoring local and cultural contexts; \textbf{Temporal Dislocation} (20.0\%) -- mis-understanding about the temporal information.}
\label{tab:error_analysis}
\end{table}
\noindent \textbf{Platform Mis-Adaptation.} In the TikTok subscriptions case, the model applies a generic economic lens, predicting high tendency toward worth option based on the influencer economy. This reasoning ignores the context of Reddit, where the user tend to be negative of influencer monetization. The prediction is thus inverted from the ground truth which shows a much higher proportion of not worth it.

\noindent \textbf{Cultural Misunderstanding.} In the phone update case, models systematically misinterpret Chinese users' pragmatic spending attitudes, failing to capture the genuine cost-conscious preferences common among Weibo users, which often contradict Western-centric consumer marketing narratives. While model recognizes about the technical enthusiasm in Weibo, their predictions fail to reflect the cost-conscious approach that characterizes technology adoption for many Chinese social media users. 

\noindent \textbf{Temporal Dislocation.} In the Twitter-to-X rebranding case, the model assumes that users would quickly adopt the new term. However, this prediction overlooks that Reddit users, in particular, demonstrated strong attachment to the familiar terminology, viewing continued use of the old name as both habitual behavior and subtle resistance. The model's temporal reasoning thus misaligned with the gradual and reluctant pace of real-world language adoption.

\section{Conclusion}
\label{sec:conclusion}
We introduce MindVote, the first benchmark for public opinion distribution prediction in social media. Our comprehensive evaluation demonstrates the critical importance of assessing models in naturalistic, context-rich environments. We argue that the path to socially intelligent AI requires enhancing a model's capacity for in-context reasoning. Our results demonstrate that models perform best when they can explicitly identify, weigh, and interpret the social cues present in the immediate context—a skill that requires flexible reasoning rather than memorized associations. MindVote provides the essential tool to guide and measure this necessary shift.

\section{Limitation}
While MindVote represents a significant advance in ecological validity, our benchmark is limited to two platforms, Reddit and Weibo, which, despite their scale and cultural diversity, do not capture the full spectrum of social media ecosystems. Platforms such as Twitter/X, TikTok, Instagram, and region-specific networks exhibit distinct interaction patterns, demographic compositions, and content modalities that may influence opinion formation differently. Future work should expand to additional platforms to ensure comprehensive evaluation of LLMs' social reasoning capabilities across diverse digital environments.
\bibliography{mindvote/aaai2026}

\end{document}